\documentclass[twocolumn]{aastex631}
\usepackage{amsmath}
\usepackage{booktabs}
\usepackage{multirow}
\usepackage{afterpage}

\begin{document}

\title{Sun-as-a-star observations of obscuration dimmings caused by filament eruptions}

\author{Yu Xu}
\affiliation{School of Earth and Space Sciences, Peking University, Beijing 100871, China, huitian@pku.edu.cn}
\affiliation{Leibniz Institute for Astrophysics Potsdam, An der Sternwarte 16, D-14482 Potsdam, Germany}
\author{Hui Tian}
\affiliation{School of Earth and Space Sciences, Peking University, Beijing 100871, China, huitian@pku.edu.cn}
\author{Astrid M. Veronig}
\affiliation{Institute of Physics, University of Graz, Graz, Austria.}
\affiliation{Kanzelhöhe Observatory for Solar and Environmental Research, University of Graz, Treffen, Austria}
\author{Karin Dissauer}
\affiliation{Institute of Physics, University of Graz, Graz, Austria.}

\begin{abstract}
Filament eruptions often lead to coronal mass ejections (CMEs) on the Sun and are one of the most energetic eruptive phenomena in the atmospheres of other late-type stars. However, the detection of filament eruptions and CMEs on stars beyond the solar system is challenging. Here we present six filament eruption cases on the Sun and show that filament material obscuring part of the solar disk can cause detectable dimming signatures in sun-as-a-star flux curves of He\,{\sc{ii}} 304 \AA. Those filament eruptions have similar morphological features, originating from small filaments inside active regions and subsequently strongly expanding to obscure large areas of the solar disk or the bright flare regions. We have tracked the detailed evolution of six obscuration dimmings and estimated the dimming properties, such as dimming depths, dimming areas, and duration. {The largest dimming depth among the six events under study is 6.2\% accompanied by the largest dimming area of 5.6\% of the solar disk area. Other events have maximum dimming depths in a range of around 1\% to 3\% with maximum areas varying between about 3\% to 4\% of the solar disk area. The duration of the dimming spans from around 0.4 hours to 7.0 hours for the six events under study.} A positive correlation was found between the dimming depth and area, which may help to set constraint on the filament sizes in stellar observations.
\end{abstract}

\section{Introduction}

Coronal mass ejections (CMEs) often result from filament eruptions and are one of the most energetic manifestation of solar magnetic activities. The occurrence of solar CMEs can alter the interplanetary environment and cause severe space weather disturbances that affect the Earth's magnetic field and atmosphere. Similar to the Sun, stars beyond the solar system {also exhibit eruptions associated with the magnetic field} and are able to launch stellar CMEs into their interplanetary spaces. Stellar CMEs bring changes to the atmospheres of exoplanets orbiting the stars and therefore affect the planetary habitability (e.g., \citealt{Lammer2007,Airapetian2020})

On the Sun, CMEs are usually observed by white-light coronagraphs which block out the much brigher emission from the solar photosphere. Detecting stellar CMEs is challenging due to the large distances to the stellar systems. The stars and their surrounding interplanetary spaces are too far and faint to be spatially resolved by current instruments. Current instruments are able to obtain the spatially integrated spectra of stars in X-ray or ultraviolet (UV) bands which are sensitive to the stellar coronal activities. Thus, attentions have been paid on searching for CME indicators standing out in the spatially integrated stellar UV and X-ray spectra (see, e.g., reviews by \citealt{Leitzinger2022,Osten2023}).

A wealth of observations and rich knowledge of solar CMEs make the sun-as-a-star observations of solar CMEs a great benchmark for studying the indicators of stellar CMEs using the spatially integrated spectra. The sun-as-a-star data can be obtained directly from observations of spectrographs such as the Extreme Ultraviolet Variability Experiment (EVE; \citealt{Woods2012}) aboard the Solar Dynamics Observatory (SDO; \citealt{Pesnell2012}) or be estimated from imaging observations by adding up the emission of each pixel. The main CME indicator in the sun-as-a-star observations at extreme-ultraviolet (EUV) wavelengths is the coronal dimming (e.g., \citealt{Thompson1998,Mason2014,Mason2016}). The term \emph{dimming} often refers to the decrease of emission in a certain waveband or the flux of a certain spectral line with respect to its quiescent level measured when no activity is detected \citep{Dissauer2019,Veronig2019}. Coronal dimmings from the Sun are regularly observed by EUV or soft X-ray (SXR) imagers to accompany the early phase of CMEs (e.g., \citealt{Hudson1996,Sterling1997,Reinard2008,Dissauer2018}). They are mostly interpreted to be caused by the depletion of plasma density resulting from the escape of the CME material (e.g., \citealt{Hudson1996,Thompson1998,Zhukov2004,Jin2009,Tian2012,Vanninathan2018}). In recent studies, coronal dimmings have been detected also on other stars in X-ray as well as EUV and far-ultraviolet (FUV) and they were regarded as an indicator of the occurrence of stellar CMEs \citep{Veronig2021,Loyd2022}.

{Another CME indicator observed in sun-as-a-star observations is the Doppler shifts of EUV spectral lines (e.g., \citealt{Xu2022,Lu2023,Otsu2024}). Previous studies have identified Doppler shifts in EUV lines resulting from plasma motions along the line-of-sight (LOS) on the Sun (e.g., \citealt{Achour1995,Brekke1997,Harra2001,Tian2012}). Furthermore, in sun-as-a-star observations, Doppler shifts associated with plasma motion during the flare process have also been reported (e.g., \citealt{Hudson2011,Brown2016,Chamberlin2016,Cheng2019}).} Similar asymmetries or Doppler shifts in spectral line profiles have also been found on other stars in the X-ray, FUV and optical bands, and were interpreted as possible {indicators} of stellar CMEs or filament eruptions (e.g., \citealt{Leitzinger2011,Vida2019,Argiroffi2019,Namekata2022,Chen2022,Lu2022,Inoue2024}).

Eruptions of solar filaments, which can develop into CMEs, often leave obvious dimming signatures in imaging observations. {Distinct from plasma evacuation,} this could be called obscuration dimming \citep{Mason2014}, which normally refers to the phenomenon that an erupting filament travels across the solar disk absorbing the background emission either from the active regions or from the quiescent Sun, and by this process causing a decrease of emission \citep{Mason2014}. \cite{Gopalswamy2013} reported an obscuration dimming in microwave caused by a prominence eruption, and derived the prominence temperature using the dimming feature. The filament obscuration in EUV was also mentioned but without detailed discussion due to the low temporal cadence of the data. Obscuration dimmings are also observed in EUV, which are believed to be caused by photonionization of neutral hydrogen and helium or singly ionized helium in the cool filament material (e.g., \citealt{Williams2013,Jenkins2018}). For instance, solar observations often reveal obscuration dimmings in narrow-band images of He\,{\sc{ii}} 304 \AA~during filament eruptions. However, such dimming signatures have been rarely investigated in detail, particularly their temporal evolution and in terms of a Sun-as-a-star analysis.

In this study, we focus on obscuration dimming caused by filament eruptions as observed in He\,{\sc{ii}} 304 \AA~. We show that filament obscurations are able to cause dimmings in the full-Sun He\,{\sc{ii}} 304 \AA~spectra and thus may serve as a proxy of the detection of stellar filament eruptions. The paper is arranged in the following manner: Section \ref{sec:observations} introduces the dataset we used and the basic properties of the filament events we analysed, Section \ref{sec:method} explains the methodology of the analysis, Section \ref{sec:results} presents the dimming properties (such as duration, dimming depth, etc.), and Sec. \ref{sec:discussion} discusses the relationship between the dimming depth and area. In Sec. \ref{sec:summary}, we summarize the main findings and present our conclusions.

\section{Instruments and Observations}\label{sec:observations}

We use 304~\AA~images captured by the Atmospheric Imaging Assembly (AIA; \citealt{Lemen2012}) onboard SDO to show the propagation of the filament material on the solar disk. SDO/AIA is an imager capturing the Sun in seven EUV wavebands (i.e. 94~\AA, 131~\AA, 171~\AA, 193~\AA, 211~\AA, 304~\AA, 335~\AA) with a time cadence of around 12 seconds and a spatial resolution of around $1.5''$ \citep{Lemen2012}. The 304~\AA~waveband has a response function that peaks at the temperature of around 0.05 MK, and is often used to image the cool filament or prominence material. We used the SDO/AIA 304~\AA~images with a time cadence of 1 minute. The images were also used to derive disk-integrated brightness in the 304~\AA~waveband.

Sun-as-a-star fluxes of the He\,{\sc{ii}} 304 \AA~line observed by SDO/EVE and the Extreme Ultraviolet Sensor (EUVS; \citealt{Eparvier2009,Snow2009}) onboard the Geostationary Operational Environmental Satellite (GOES-R) were adopted to track the temporal evolution of the line intensities during filament eruptions. The EUV spectrograph onboard SDO has two instruments inside (i.e. MEGS-A and MEGS-B). They provide sun-as-a-star spectra ranging from 3 nm to 103 nm with a temporal resolution of 10 seconds and a spectral resolution of 1 \AA~\citep{Woods2012}. The short wavelength part of the spectra (i.e. from 3 nm to 37 nm) is unavailable since 2014 May due to an anomaly of MEGS-A and the time cadence of the MEGS-B spectra changes to 1 minute afterwards to maintain a high signal to noise ratio. The GOES-R/EUVS instrument observes several distinct solar emission lines including the He\,{\sc{ii}} \AA~line with a temporal resolution of 1 minute. We generated flux curves of the He\,{\sc{ii}} 304 \AA~line as a function of time based on the data from SDO/EVE and GOES-R/EUVS. The time cadence in our analysis is 1 minute.

The obscuration dimming events were identified by cross-matching two catalogs. The EVE flare catalog \footnote{https://lasp.colorado.edu/eve/data\_access/eve-flare-catalog/index.html} \citep{Hock2012} records flare events from 2010 May 1 to 2014 November 21. The flare catalog contains plots that provide a quick look of the temporal evolution of selected line intensities observed by SDO/EVE during the flare events. The filament catalog \footnote{https://aia.cfa.harvard.edu/filament/} \citep{McCauley2015} contains information of 1004 filament eruptions from 2010 May 18 to 2014 September 20. We searched through these two catalogs and found four filament eruptions showing filament obscuration on the disk as well as dimmings in the intensity evolution of the He\,{\sc{ii}} 304 \AA~line. The four events with their flare information are listed in Table \ref{tab:cases}. The flare information is adopted from the GOES event list \footnote{https://hesperia.gsfc.nasa.gov/goes/goes\_event\_listings} and {the corresponding CME speeds are adopted from LASCO CME list \footnote{https://cdaw.gsfc.nasa.gov/CME\_list/}}. The other two events (i.e. Case 5 and Case 6) in Table \ref{tab:cases} were found by manually monitoring large filament eruptions in the current solar cycle.

\section{Method}\label{sec:method}
\subsection{Identification of Obscured Regions}\label{sec:identification}
A time series of AIA 304 \AA~images were used to identify the regions obscured by the propagating filaments and to trace the temporal evolution of the areas of the obscured regions. {The method involves setting a threshold value of the emission in order to separate the dark regions associated with the propagating filaments from other regions on the solar disk, as detailed below.} 

The time series of the AIA 304~\AA~images starts from at least 30 minutes before the flare onset and spans the whole occurrence of the obscuration dimming with a temporal resolution of 1 minute. We refer to the first 30 minutes of the dataset when no obvious solar activities are detected to be the \emph{pre-flare period} (as listed in Table \ref{tab:cases}). A reference 304 \AA~image was created with flux at each pixel calculated by taking the average pixel values of the images obtained during the pre-flare period. This pre-flare 304~\AA~image was used to determine the threshold value. 

\begin{figure*}
    \centering
    \includegraphics[width=1.0\textwidth]{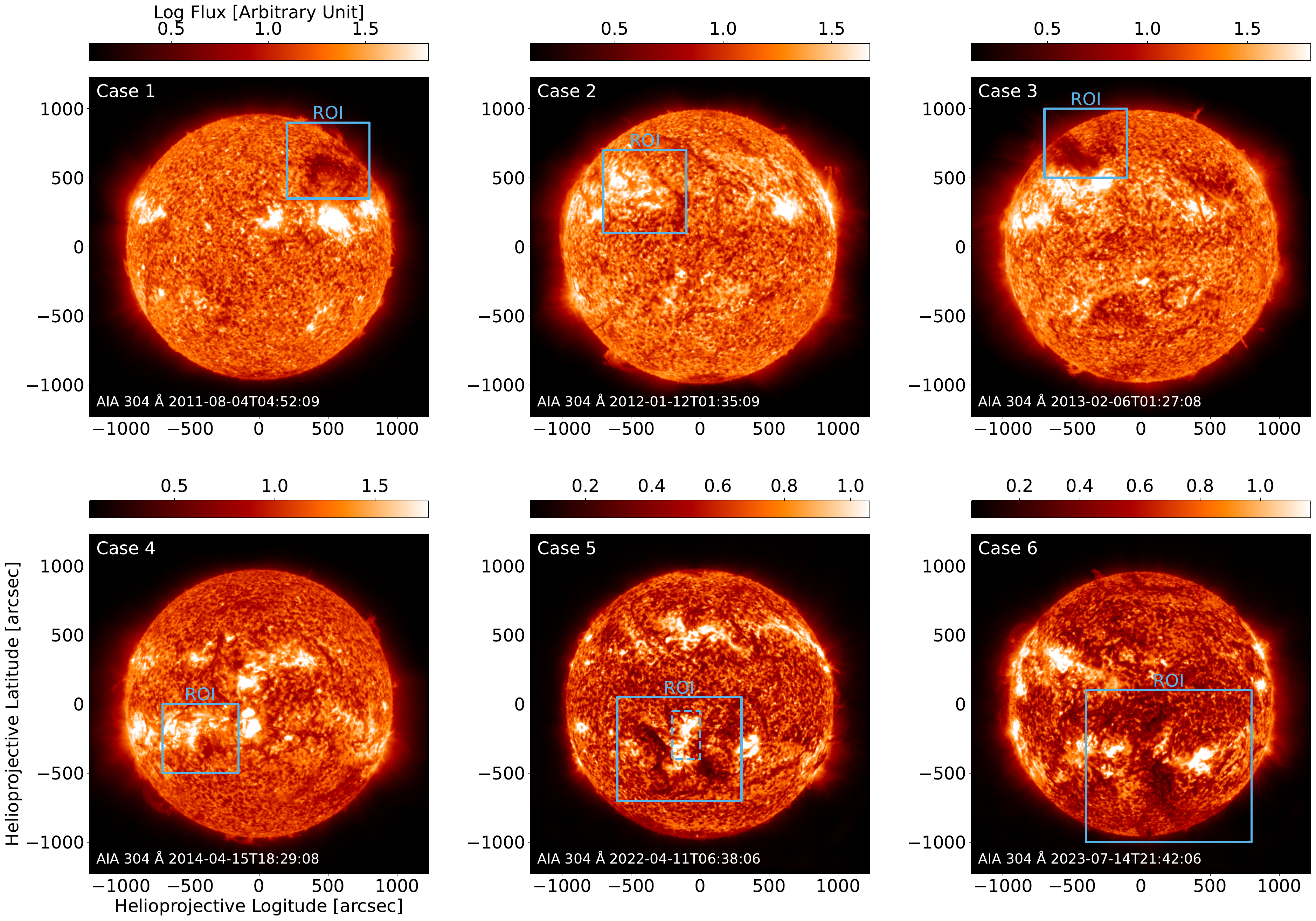}
    \caption{Overview on the six filament obscuration dimmings and regions of interest (ROIs) outlined on AIA 304~\AA~images by the blue solid lines.  The blue dashed lines in Case 5 enclose the flare region which is excluded from the ROI.}
    \label{fig:roi}
\end{figure*}

For each event, we defined a region of interest (ROI) which includes the dimming regions during their whole evolution. In Fig. \ref{fig:roi}, we outline the ROI in the six events under study using blue solid lines to show two types of ROI. The ROIs in Case 1--4 and 6 are rectangular regions enclosed by the blue solid lines. Most parts of the flare regions in Case 1, Case 3, and Case 6 are outside the rectangular ROIs, which means that the flare regions are not obscured by the filament material. The flare regions in Case 2 and Case 4 are inside the rectangular ROIs because the erupting filaments obscure part of the flare regions in those two cases. Case 5 has a different shape of the ROI compared to other cases because the filament material in Case 5 is scattered around its flare regions and do not obscure the flare regions. In Case 5 there is another area outlined by the blue dashed line inside the blue solid rectangle, showing the region that is not obscured by the filament material during their propagation. The ROI in Case 5 refers to the region within the solid rectangle but outside the dashed rectangle. The threshold value is set to the median value of the pixel fluxes in the ROI on the pre-flare image, representing the flux level during the pre-flare period. The regions including pixel values smaller than the threshold value were marked as dark regions. For high calculation efficiency, we only let the ROI instead of the whole AIA 304 \AA~image pass through the identification procedure.

\begin{figure*}
    \centering
    \includegraphics[width=1.0\textwidth]{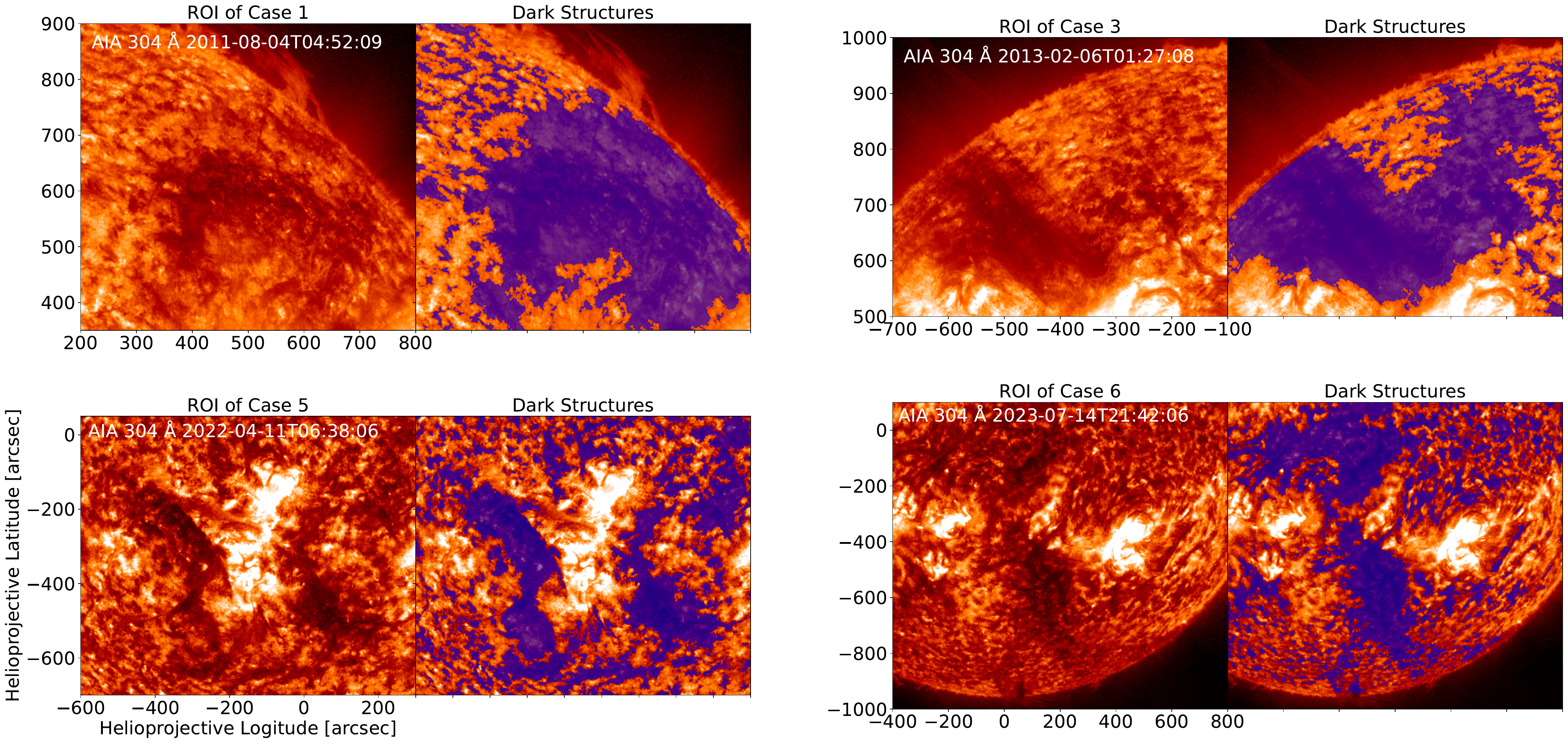}
    \caption{Identification results of the obscured regions in Case 1, Case 3, Case 5, and Case 6 at the time step when the total region area reaches its maximum. In each panel, the left is the original image of the ROI with the time of the snapshot labeled, whereas the blue masks on the right indicate the dark structures whose fluxes fall below the threshold value (see Sec. \ref{sec:identification}). It is worth noting that the dark structures include both the regions obscured by the erupting filaments and the dark fibrils on the solar disk (see the main text).}
    \label{fig:filament}
\end{figure*}

Figure \ref{fig:filament} shows the results of the identification from Case 1, Case 3, Case 5, and Case 6 at the time step when the total area of the obscured regions reaches its maximum value. On each panel of Fig. \ref{fig:filament}, the 304~\AA~image of the ROI is plotted on the left, whereas on the right we mask the identified dark regions by the blue color. The area of the dark regions is estimated at each time step by summing up the areas of the pixels inside the dark regions. The result is converted to the unit of the percentage of the solar disk area $\rm{S_\odot}$. It is worth noting that the identified dark regions include not only the obscured regions but also the dark fibrils on the solar disk. {In order to estimate the area of the obscuration dimming, we have to remove the contamination from the fibrils in the area estimation. Because those fibrils barely change compared to the filament during the obscuration dimmings, we estimated the areas of the fibrils using the images obtained during the pre-flare period when the identified dark regions only contain the fibrils.} We calculated the areas of the fibrils at each time step during the pre-flare period and averaged the areas over time to obtain a representative area of the fibrils. We subtracted this area of the fibrils from the area of the dark regions at each time step during the whole time series to obtain the area of the dimming regions and to study its temporal evolution.

The identification procedure does not work well for Case 2 and Case 4 because most parts of the obscured regions are inside the bright flare regions. The pixel fluxes of the obscured regions do not fall below the threshold value. Since under this conditions it is ambiguous to separate the obscured regions from other regions using the aforementioned method, we did not estimate the areas of the dimming regions in those two cases.

\subsection{Sun-as-a-star Dimmings}\label{sec:dimmingsigs}
The SDO/AIA $304$ \AA~imaging data were used to estimate the sun-as-a-star flux in the $304$~\AA~waveband. We intended to show that the filament material obscuring parts of the solar disk during its propagation can cause dimmings in the temporal evolution of the disk-integrated flux. We added up the pixel fluxes on the spatially resolved AIA 304 \AA~images and used the sum to represent the sun-as-a-star flux in the 304 \AA~waveband, denoted as $F_{total}$. We estimated the pre-flare level of the flux (denoted as $F_{total}(t_0)$) by averaging the total flux over the pre-flare period. We divided $F_{total}$ by $F_{total}(t_0)$ at each time step to obtain the relative flux.

We also analyzed the temporal evolution of the total flux inside the ROI. The total flux in the ROI (denoted as $F_{ROI}$) was calculated by adding the pixel fluxes inside the ROI. The pre-flare level $F_{ROI}(t_0)$ was estimated by averaging $F_{ROI}$ over the pre-flare period.

The properties of the obscuration dimmings in flux curves are described by the dimming depth and the dimming duration. A criterion was set to identify the dimming signature in the temporal evolution of fluxes: the flux at a certain time step (denoted as $F(t)$) drops to lower than its pre-flare level minus its standard deviation during the pre-flare period (denoted as $F(t_0)-\sigma$, where $\sigma$ is the standard deviation). The dimming depth is defined as the percentage of the decrease of the flux with respect to its pre-flare level, namely $(F(t)-F(t_0))/F(t_0)$. The duration of the dimming is the length of the time period when the flux curve shows the dimming signature. 

\section{Results}\label{sec:results}
We studied six filament eruptions showing obscuration dimming signatures in the AIA $304$~\AA~filtergram. The six eruptions share a similar behavior in that they originate from small filaments inside active regions and subsequently strongly expand to obscure broader regions of the solar disk and/or the bright flare regions. In two out of six events (Case 2 and Case 4), the filament material obscures the emission from the underlying flare regions. Three events (Case 1, Case 3, and Case 5) show a situation where the dense filament material obscures the emission from the quiescent Sun in the background. Case 6 is a special case where the filament material obscures the flare region at the very beginning of its propagation process and then extends to obscure quiescent regions in the later evolution stage.

\begin{figure*}
    \centering
    \includegraphics[width=1.0\textwidth]{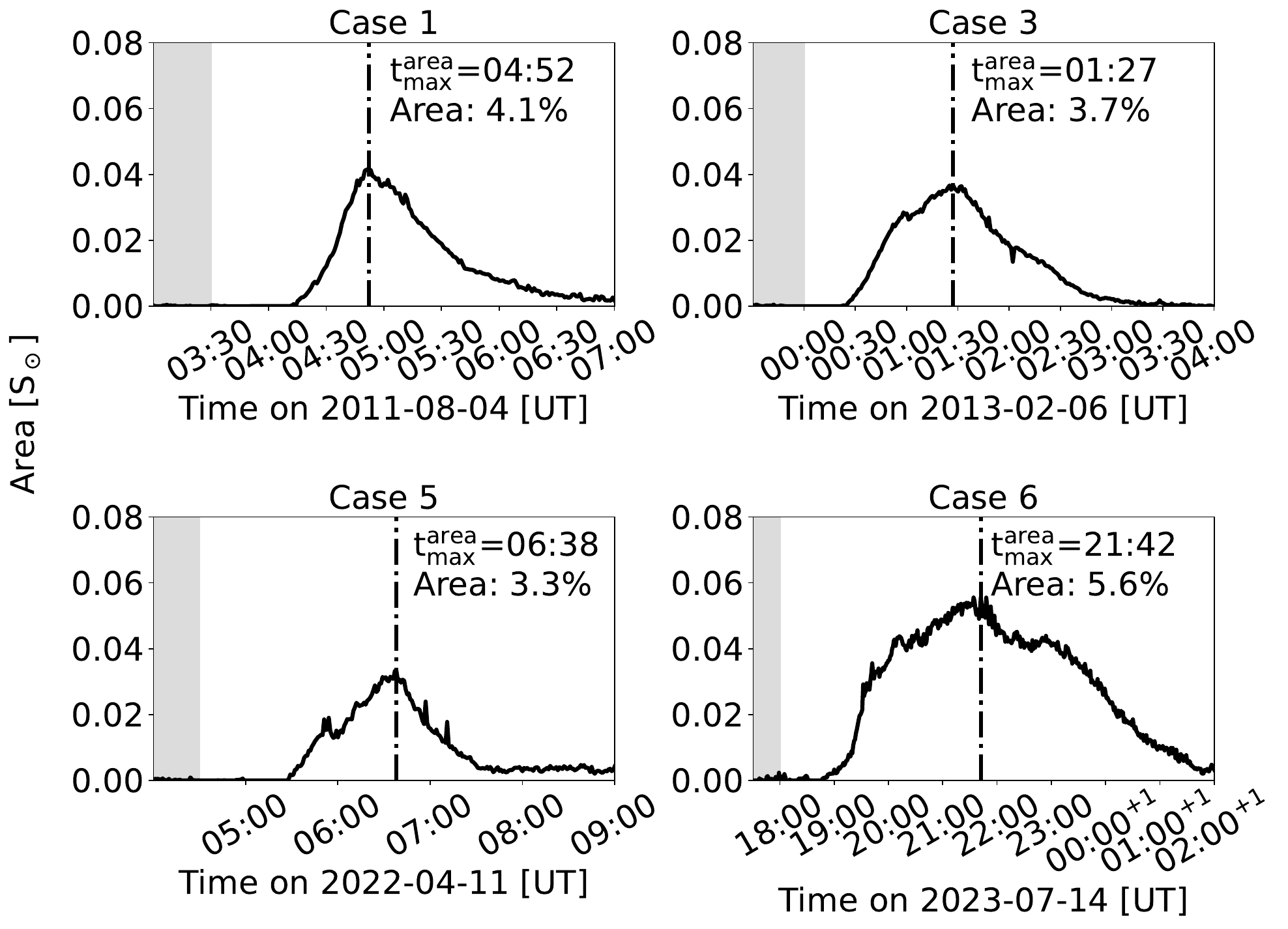}
    \caption{Temporal evolution of the areas of the dimming regions in Case 1, Case 3, Case 5, and Case 6. The grey shade indicates the pre-flare period (see Sec. \ref{sec:identification} for details). The vertical dashed line indicates the time $t_{max}^{area}$ when the area of the dimming region reaches its maximum value. The corresponding time and area are labeled on each panel. The format of the time is Hour:Minute.}
    \label{fig:filamentareas}
\end{figure*}

Figure \ref{fig:filamentareas} shows the temporal evolution of the areas of the obscuration dimmings in the four cases. The grey shade in each panel in Fig. \ref{fig:filamentareas} marks the pre-flare period when no obvious flare activities are detected on the solar disk. The vertical dashed line in each panel indicates the time ($t_{max}^{area}$) when the dimming area reaches its maximum value. The maximum area of the obscuration dimming during its evolution and its corresponding time $t_{max}^{area}$ for each case are summarized in Table \ref{tab:dimming}. {Case 6 shows the largest maximum dimming area among the six events under study, with the dimming regions covering up to 5.6\% of the solar disk area ($\rm{S_\odot}$), whereas the other cases have maximum areas in the range of about 3\% $\rm{S_\odot}$ to 4\% $\rm{S_\odot}$}.

\begin{figure*}
    \centering
    \includegraphics[width=1.0\textwidth]{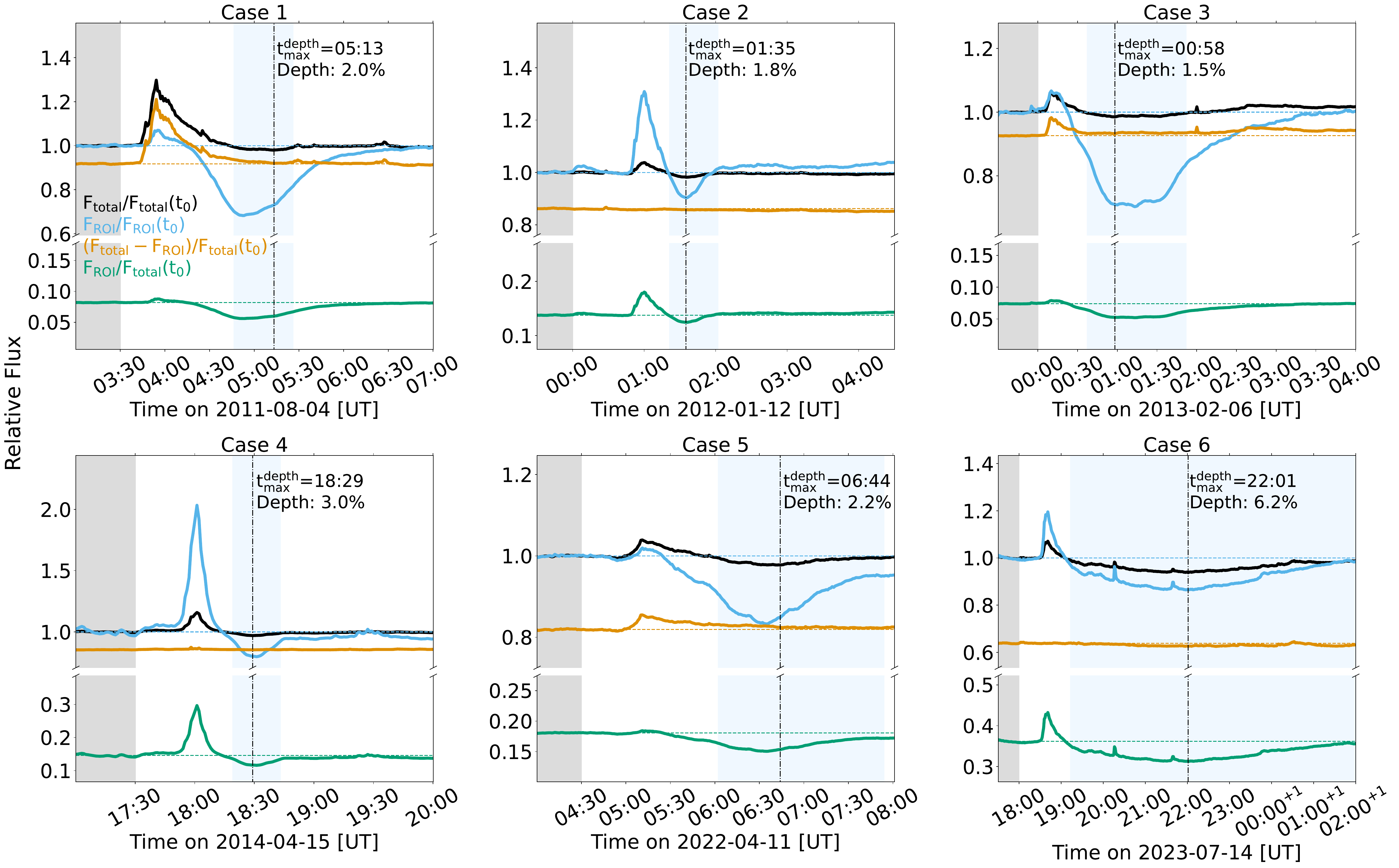}
    \caption{Temporal evolution of fluxes in the AIA 304~\AA~waveband for all six cases. The black curve represents the temporal evolution of the ratio between $F_{total}$ and $F_{total}(t_0)$ which indicates the evolution of the sun-as-a-star flux. The blue curve plots $F_{ROI}/F_{ROI}(t_0)$ which shows the temporal evolution of the flux inside the ROI with respect to its pre-flare level. The green curve marks the contribution from the ROI, which is $F_{ROI}/F_{total}(t_0)$. The orange curve is $(F_{total}-F_{ROI})/F_{total}(t_0)$, representing the contribution of the flux from the region outside the ROI. The black dashed line indicates the time $t_{max}^{depth}$ when the disk-integrated flux (i.e. $F_{total}$) reaches its minimum and the corresponding maximum dimming depth is labeled beside it. The grey shade marks the pre-flare period.}
    \label{fig:relative_flux}
\end{figure*}

In Fig. \ref{fig:relative_flux} we show the time evolution of the relative fluxes derived from the AIA 304~\AA~images for all six cases. The black curve is $F_{total}/F_{total}(t_0)$, representing the evolution of the sun-as-a-star flux with respect to its pre-flare level. All the six cases show a decrease of the flux after the corresponding flare and recovery of the flux after the filament material is barely visible on the disk. We refer to this decrease and recovery processes of the flux to be the dimming signatures. We characterized the dimmings by their duration and their dimming depths. The duration of dimming means the length of the time period when an obscuration dimming occurs (marked by the blue shades in Fig. \ref{fig:relative_flux}). The dimming depth is defined as the percentage of the decreased flux with respect to its pre-flare level. The time when the dimming depth reaches its maximum value during the evolution is denoted as $t_{max}^{depth}$ which is indicated by the vertical dashed line in Fig. \ref{fig:relative_flux}. Table \ref{tab:dimming} lists the duration, maximum dimming depth and the corresponding time ($t_{max}^{depth}$) for each of the six cases. The dimming in Case 6 lasts for the longest time among the six cases, 6.78 hours in total. Case 6 also has the largest maximum dimming depth among all the cases, reaching 6.2\% with respect to its pre-flare level. {The other cases show substantially smaller values, with a maximum dimming depth between 1.5\% to 3.0\% and a duration spanning from 0.4 hours to 1.9 hours.}

\begin{figure*}
    \centering
    \includegraphics[width=1.0\textwidth]{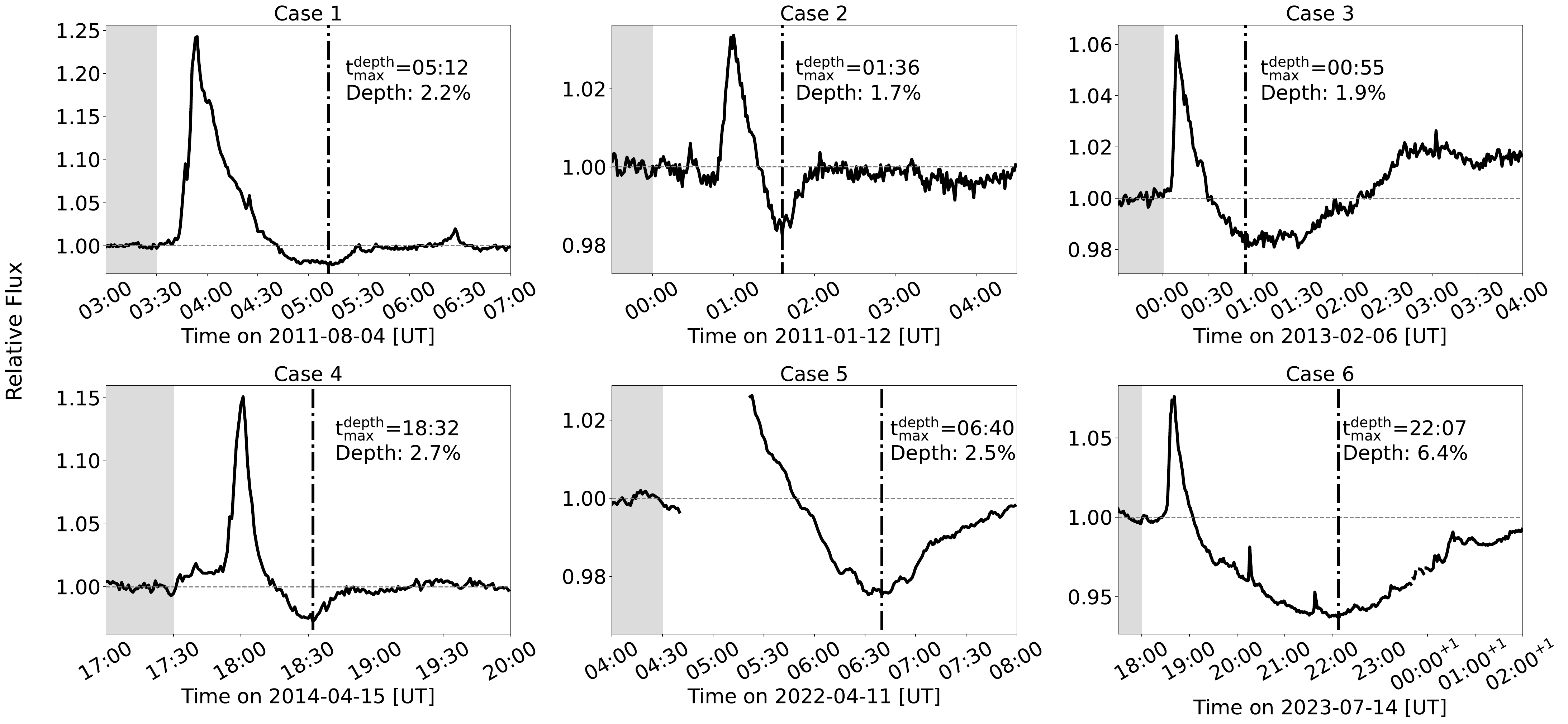}
    \caption{Temporal evolution of the line intensity of He\,{\sc{ii}} 304 \AA~from sun-as-a-star spectroscopic observations. The Case 1--4 were observed by SDO/EVE. The last two cases, Case 5 and Case 6, were observed by GOES-R/EUVS. The grey shade in each panel indicates the pre-flare period. The maximum dimming depth and its corresponding time are labeled on each panel.}
    \label{fig:eveeuvs}
\end{figure*}

{Figure \ref{fig:eveeuvs} shows the time evolution of the He II 304 \AA~line observed by SDO/EVE and GOES-R/EUVS. One can see that also in these Sun-as-a-star observations the obscuration dimming signatures are well visible for each of the six events.} The flux curves show dimmings after the flare events as well as the recovery of the dimmings. In each case, the time when the dimming reaches its maximum dimming depth is indicated by the vertical dashed line.  The maximum dimming depths as well as the corresponding times ($t_{max}^{depth}$) are labeled beside the vertical lines {and their values are listed in Table \ref{tab:dimming}.} The $t_{max}^{depth}$ and maximum dimming depths derived here agree well with those derived from the flux curves of AIA $304$~\AA~waveband. There are slight differences because the SDO/AIA 304 \AA~waveband has a wider temperature response, so the images contain contributions from the emission of plasma over a broader temperature range.

\section{Discussion}\label{sec:discussion}
Filament eruptions accompanied by obscuration dimming signatures in the sun-as-a-star observations are not commonly seen. The first four cases (Case 1--4) were selected from cross-matching two catalogs among over 1000 filament eruptions. Other filament eruptions without obscuration dimming features include eruptions originated from quiet-Sun filaments as well as active region filaments. The eruptions caused by large quiet-Sun filaments can also obscure broad regions during their propagation. However, the filaments are often elongated and slightly widened during their evolution, therefore causing only small changes to the area coverage of the filament material and almost no dimming signatures in the disk-integrated flux curves of He\,{\sc{ii}} 304 \AA. Other erupting filaments from active regions may not be dense enough to maintain high density during the expansion, so the obscured regions disappear very fast or the dimming signatures are buried in the flare enhancement. The six eruptions with obscuration dimmings in the sun-as-a-star observations found in this work all originated from small filaments inside active regions. The erupting filaments develop into dense structures covering large areas of the solar disk or the bright flare regions. The filaments are able to maintain high density for a relatively long time, so they absorb a large amount of emissions from the background and offset the flare enhancement in the flux curves.

It is worth noting that Case 6 has the largest maximum dimming depth as well as the largest maximum obsucration dimming area. {This directs the investigation into the correlation between the dimming depth and area.} For the four cases with area estimation (i.e., Case 1, Case 3, Case 5, and Case 6), we estimated the dimming depth and dimming area at each time step during the dimming period. The dimming depth was estimated from the flux curve of the AIA 304 \AA~filtergram and the dimming area was estimated based on the identification outputs described in Sec. \ref{sec:identification}. 

\begin{figure*}
    \centering
    \includegraphics[width=1.0\textwidth]{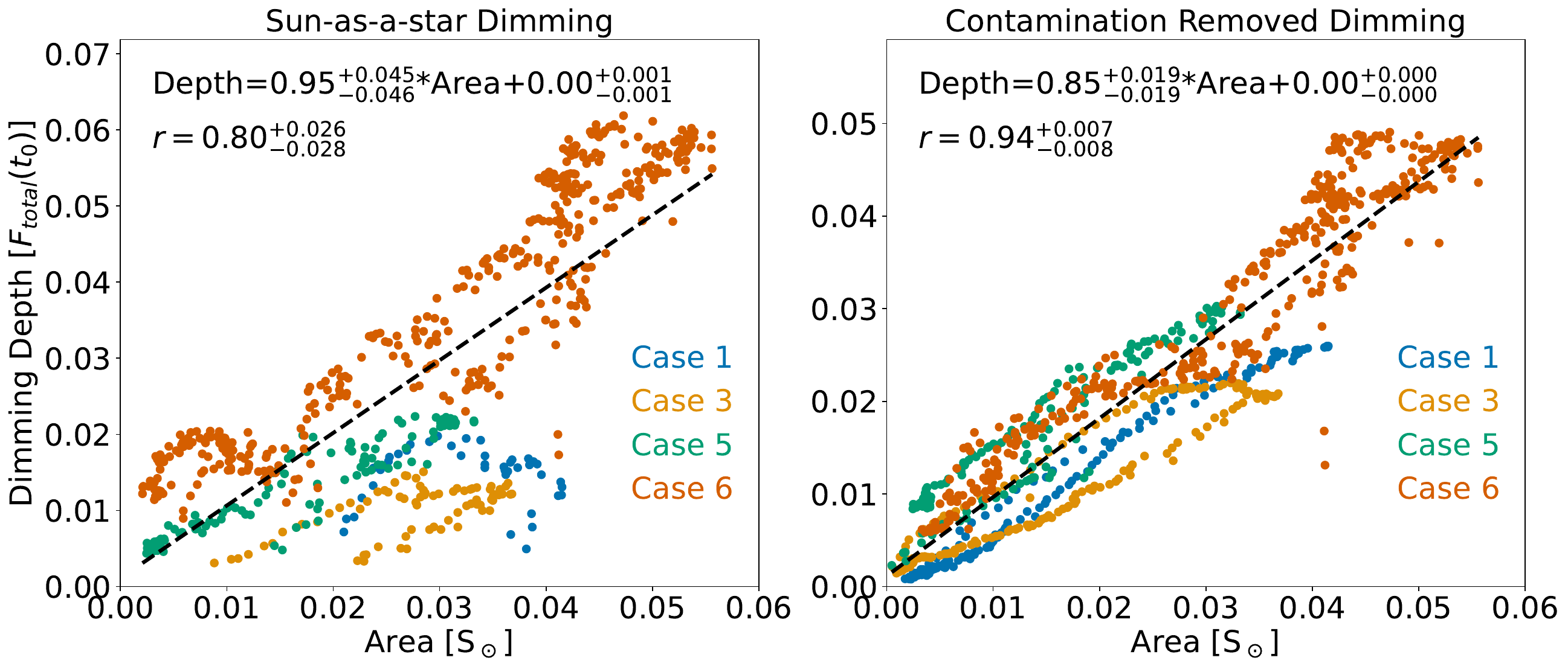}
    \caption{The relationship between the dimming area and the dimming depth based on the data of 4 cases with area estimation. Different colors mark the different cases. The area has a unit of the percentage of the solar disk area $\rm{S_\odot}$. In the left panel the dimming depths were estimated from the sun-as-a-star flux curves derived from SDO/AIA 304 \AA~images. In the right panel, the dimming depths were calculated after we removed the contributions from the regions outside the main dimming regions (i.e., ROI). Linear fitting was conducted and the fitted lines are plotted as the black dashed lines. The corresponding Pearson Correlation Coefficient $r$ and fitting equation are labeled on each panel. {The ranges of the fitting parameters and $r$ represent 95\% confidence intervals obtained by applying the bootstrapping technique for 10000 iterations.}}
    \label{fig:areadepth}
\end{figure*}

In Fig. \ref{fig:areadepth} (left panel), we plotted the dimming depths against the dimming areas obtained from Case 1, Case 3, Case 5 and Case 6, showing a positive correlation, but with different slopes for the four cases. The correlation is best developed for Case 6, which has the largest range in the dimming area and depth over the time evolution. We conducted a linear fitting to the scatter plot and the fitted line is plotted as the dashed black line. {The Pearson Correlation Coefficient $r$ of the fitting is 0.80 , indicating a strong correlation between the dimming area and depth. The ranges of the fitting parameters and $r$ (labeled on Fig. \ref{fig:areadepth}) were obtained from the bootstrapping technique of 10000 iterations and represent the 95\% confidence intervals.} It is worth mentioning that there are points which are obviously deviating from the linear fitting results. Also, there is a conspicuous time difference between $t_{max}^{area}$ and $t_{max}^{depth}$ (see Table \ref{tab:dimming}), meaning that the dimming area and the dimming depth do not reach their maximum values simultaneously, which challenges the positive correlation between the dimming area and depth. 

We suggest that the existence of the straying points on the left panel of Fig. \ref{fig:areadepth} could be attributed to (1) the contribution of fluxes from the flare regions affecting the estimation of dimming depths, and (2) the uncertainties in the area estimation and the flux measurement. The contamination from the flare regions and other parts of the Sun is shown by the orange curves in Fig. \ref{fig:relative_flux}. The orange curve in each case represents the flux outside the ROI, i.e. outside the region where the obscuration dimming mainly occurs. During the dimming period (marked by the blue shade in each panel), the orange curve does not preserve its pre-flare value, demonstrating that the emissions from outside the ROI contaminate the dimming depth estimation. We removed the contamination by summing up solely the emissions inside the ROI. The ROI flux curve is plotted as the green curve in Fig. \ref{fig:relative_flux} { and it is scaled to its pre-flare value. The dimmings start earlier and have generally larger depths in the ROI curves than those in the sun-as-a-star curves. We estimated the dimming depths from the ROI curves at each time step during their dimming periods and plotted the dimming depths against the dimming areas on the right panel of Fig. \ref{fig:areadepth}.} After removing the contamination from the emissions of regions outside ROIs, the positive correlation still holds and becomes tighter than that in the left panel. The Pearson Correlation Coefficient $r$ rises to 0.94 {with a 95\% confidence interval between 0.93 and 0.95}, indicating a significant correlation between the dimming depth and area.

{We can derive the linear correlation theoretically. We denote the total area of dimming regions as $S_{d}$ and the depth of dimmings as $D$. We use $f_{d}$ and $f_{nd}$ to represent the area averaged emission level inside and outside the dimming region, respectively. Time $t_0$ represents the pre-flare period and $t$ is some dimming time. We introduce the emission level averaged over the disk area $S_\odot$ in the pre-flare period as $f(t_0)$, which gives
\begin{equation}
f(t_0)S_\odot=f_d(t_0)S_d+f_{nd}(t_0)(S_\odot-S_d)
\label{fto}
\end{equation}
The dimming depth at time $t$ can be written as,
\begin{equation}
\begin{split}
&D(t)=\frac{\big[f_d(t_0)-f_d(t)\big]S_d+\big[f_{nd}(t_0)-f_{nd}(t)\big](S_\odot-S_d)}{f(t_0)S_\odot}\\
& = \frac{f_d(t_0)-f_d(t)+f_{nd}(t)-f_{nd}(t_0)}{f(t_0)}\frac{S_d}{S_\odot}+\frac{f_{nd}(t_0)-f_{nd}(t)}{f(t_0)}
\end{split}
\label{dt}
\end{equation}}

{We first assume that there is no significant activity in the non-dimming region so its average emission level does not change so much. We will have $f_{nd}(t_0)\sim f_{nd}(t)$ and the intercept in Eq. \eqref{dt} will approach 0. The slope in Eq. \eqref{dt} becomes $\frac{f_d(t_0)-f_d(t)}{f(t_0)}$. We can further assume that the filament material is dense enough to absorb all the emission from the background, which yields $\frac{f_d(t)}{f(t_0)}\sim 0$. Eq. \eqref{dt} can be written as, 
\begin{equation}
D(t)=\frac{f_d(t_0)}{f(t_0)}\frac{S_d}{S_\odot}
\end{equation}
If during the pre-flare period the flux is homogeneously distributed over the disk, we will have $f(t_0)\sim f_d(t_0)\sim f_{nd} (t_0)$ and the slope of Eq. \eqref{dt} will approach 1.}

{Our fitting results in Fig. \ref{fig:areadepth} show intercepts close to 0, which agrees well with the fact that there is no significant activity outside the dimming region. The fitted slopes approach 1 but a bit smaller than 1 because the pre-flare flux is not evenly distributed due to the existence of several active regions.}

{The establishment of the linear correlation could be useful in estimating the dimming area when only the disk-integrated flux curves are available in the observations. This is usually the case for stellar observations. Under the assumptions that (1) there is no activities outside the dimming region, (2) the emission level in the dimming region is low compared to its pre-flare emission level, and (3) the flux is homogeneously distributed over the stellar disk, the correlation between the dimming depth and area can be written as,
\begin{equation}
D(t)=\frac{S_d}{S_\bigstar}
\end{equation}
where $S_\bigstar$ is the stellar disk area. This correlation may help to constrain the size of the stellar filament using the dimming signatures in the flux curves of He\,{\sc{ii}} 304 \AA~.}

\section{Conclusions}\label{sec:summary}
We found six filament eruptions on the Sun with obscuration dimming signatures in sun-as-a-star He\,{\sc{ii}} 304 \AA~narrow-band imaging and full-Sun spectroscopic observations, out of a total of about 1000 events that were inspected. This indicates that solar filament obscuration dimmings are rare events but can can be detected in the Sun-as-a-star observations. The six eruptions all originate from small filaments inside active regions and then expand to obscure broader regions of the solar disk. These results suggest that the obscuration dimming signature is promising to serve as an indicator of stellar filament eruptions.

{The expansion and dissipation of the dimming regions are reflected by the variation of the dimming areas. The maximum dimming area among the six events is 5.6\% $\rm{S_\odot}$ and other cases have area coverage between around 3\% $\rm{S_\odot}$ to 4\% $\rm{S_\odot}$ . The dimming signatures in the full-Sun flux curves are characterised by the dimming depths and dimming duration. The maximum dimming depths of the six cases range from around 1\% to 6\% with respect to their corresponding full-Sun fluxes in 304 \AA~waveband during the pre-flare periods. The duration of the dimmings spans from 0.4 hours to a maximum of 6.8 hours.}

{A positive correlation between the dimming depth and dimming area was found with a Pearson Correlation Coefficient $r$ of $0.80^{+0.026}_{-0.028}$. The correlation becomes tighter ($r=0.94^{+0.007}_{-0.008}$) after removing the contamination from regions outside the main dimming regions. {If similar obscuration dimming signatures can be found in stellar observations, and the dimming depths can be estimated accordingly, then this correlation has the potential to facilitate setting constraints for estimations of the sizes of erupting stellar filaments.}}

\begin{longrotatetable}
\begin{table*}
    \centering
    \caption{Six cases of obscuration dimmings and their flare/CME information}
    \begin{tabular}{*6{c}}
        \hline
        \hline
        \multirow{2}{*}{Case Id}  & Pre-flare Period & Flare Onset  & Flare Peak &\multirow{2}{*}{Flare Class}  & CME Speed\\
        &  [UT] & [UT] & [UT] & &[km/s]\\
         \hline
        1 & 2011.08.04 03:00 -- 2011.08.04 03:30 & 2011.08.04 03:41 & 2011.08.04 03:57 & M9.3 & 1315\\
        2 & 2012.01.11 23:30 -- 2012.01.12 00:00 & 2012.01.12 00:49 & 2012.01.12 00:58 & C1.5 & 211\\
        3 & 2013.02.05 23:30 -- 2013.02.06 00:00 & 2013.02.06 00:04 & 2013.02.06 00:21 & C8.7 & 1867\\
        4 & 2014.04.15 17:00 -- 2014.04.15 17:30 & 2014.04.15 17:53 & 2014.04.15 17:59 & C7.3 & 360\\
        5 & 2022.04.11 04:00 -- 2022.04.11 04:30 & 2022.04.11 04:59 & 2022.04.11 05:21 & C1.6 & 940\\
        6 & 2023.07.14 17:30 -- 2023.07.14 18:00 & 2023.07.14 18:27 & 2023.07.14 18:44 & C8.8 & 862\\
        \hline
    \end{tabular}
    \label{tab:cases}
\end{table*}

\begin{table*}
    \centering
    \caption{Properties of obscuration dimmings in six cases}
    \begin{tabular}{*9{c}}
        \hline
        \hline
        \multirow{3}{*}{Case Id} & \multirow{2}{*}{Start Time} &  \multirow{2}{*}{Duration} & \multirow{2}{*}{Maximum Area}  & \multirow{2}{*}{$t_{max}^{area}$} & \multicolumn{2}{c}{AIA} & \multicolumn{2}{c}{EVE/EUVS}\\
        \cmidrule{6-7} \cmidrule{8-9}
         &&&&&Maximum Depth  & $t_{max}^{depth}$ & Maximum Depth  & $t_{max}^{depth}$ \\
         & [UT] & [hr] & [$\rm{S_\odot}$] & [UT] & [$\rm{F_{total}(t_0)}$] & [UT] & [$\rm{F_{total}(t_0)}$] & [UT]\\
         \hline
        1 & 2011.08.04 04:46 & 0.67 &4.1\%& 2011.08.04 04:52 & 2.0\%& 2011.08.04 05:13& 2.2\%& 2011.08.04 05:12 \\
        2 & 2012.01.12 01:21 & 0.68 &N/A  & N/A &1.8\%& 2012.01.12 01:35 & 1.7\%& 2012.01.12 01:36\\
        3 & 2013.02.06 00:37 & 1.25 &3.7\%& 2013.02.06 01:27 &1.5\%& 2013.02.06 00:58& 1.9\% & 2013.02.06 00:55 \\
        4 & 2014.04.15 18:19 & 0.40 &N/A  & N/A &3.0\%& 2014.04.15 18:29 & 2.7\% & 2014.04.15 18:32 \\
        5 & 2022.04.11 06:02 & 1.87 &3.3\%& 2022.04.11 06:38 &2.2\%& 2022.04.11 06:44 & 2.5\% &2022.04.11 06:40 \\
        6 & 2023.07.14 19:13 & 6.78 &5.6\%& 2023.07.14 21:42 &6.2\%& 2023.07.14 22:01& 6.4\%& 2023.07.14 22:07 \\
        \hline
    \end{tabular}
    \label{tab:dimming}
\end{table*}
\end{longrotatetable}

\section*{Acknowledgments}
This work was supported by the National Natural Science Foundation of China grants 12250006, 12425301, and the New Cornerstone Science Foundation through the XPLORER PRIZE. AMV acknowledges the Austrian Science Fund (FWF) under the grant 10.55776/14555. We acknowledge support by ISSI to the team "Coronal dimmings and their relevance for the physics of coronal mass ejections". We would like to acknowledge the data use from \emph{GOES-R} and \emph{SDO}. SDO is a mission of NASA’s Living With a Star Program. The CME catalog is generated and maintained at the CDAW Data Center by NASA and The Catholic University of America in cooperation with the Naval Research Laboratory. SOHO is a project of international cooperation between ESA and NASA. Y. X. is funded by China Scholarship Council to visit AIP.

\bibliographystyle{aasjournal}
\bibliography{ref}{}

\newpage
\begin{thebibliography}{}
\expandafter\ifx\csname natexlab\endcsname\relax\def\natexlab#1{#1}\fi
\providecommand{\url}[1]{\href{#1}{#1}}
\providecommand{\dodoi}[1]{doi:~\href{http://doi.org/#1}{\nolinkurl{#1}}}
\providecommand{\doeprint}[1]{\href{http://ascl.net/#1}{\nolinkurl{http://ascl.net/#1}}}
\providecommand{\doarXiv}[1]{\href{https://arxiv.org/abs/#1}{\nolinkurl{https://arxiv.org/abs/#1}}}

\bibitem[{{Achour} {et~al.}(1995){Achour}, {Brekke}, {Kjeldseth-Moe}, \& {Maltby}}]{Achour1995}
{Achour}, H., {Brekke}, P., {Kjeldseth-Moe}, O., \& {Maltby}, P. 1995, \apj, 453, 945, \dodoi{10.1086/176454}

\bibitem[{{Airapetian} {et~al.}(2020){Airapetian}, {Barnes}, {Cohen}, {Collinson}, {Danchi}, {Dong}, {Del Genio}, {France}, {Garcia-Sage}, {Glocer}, {Gopalswamy}, {Grenfell}, {Gronoff}, {G{\"u}del}, {Herbst}, {Henning}, {Jackman}, {Jin}, {Johnstone}, {Kaltenegger}, {Kay}, {Kobayashi}, {Kuang}, {Li}, {Lynch}, {L{\"u}ftinger}, {Luhmann}, {Maehara}, {Mlynczak}, {Notsu}, {Osten}, {Ramirez}, {Rugheimer}, {Scheucher}, {Schlieder}, {Shibata}, {Sousa-Silva}, {Stamenkovi{\'c}}, {Strangeway}, {Usmanov}, {Vergados}, {Verkhoglyadova}, {Vidotto}, {Voytek}, {Way}, {Zank}, \& {Yamashiki}}]{Airapetian2020}
{Airapetian}, V.~S., {Barnes}, R., {Cohen}, O., {et~al.} 2020, International Journal of Astrobiology, 19, 136, \dodoi{10.1017/S1473550419000132}

\bibitem[{{Argiroffi} {et~al.}(2019){Argiroffi}, {Reale}, {Drake}, {Ciaravella}, {Testa}, {Bonito}, {Miceli}, {Orlando}, \& {Peres}}]{Argiroffi2019}
{Argiroffi}, C., {Reale}, F., {Drake}, J.~J., {et~al.} 2019, Nature Astronomy, 3, 742, \dodoi{10.1038/s41550-019-0781-4}

\bibitem[{{Brekke} {et~al.}(1997){Brekke}, {Kjeldseth-Moe}, \& {Harrison}}]{Brekke1997}
{Brekke}, P., {Kjeldseth-Moe}, O., \& {Harrison}, R.~A. 1997, \solphys, 175, 511, \dodoi{10.1023/A:1004950330900}

\bibitem[{{Brown} {et~al.}(2016){Brown}, {Fletcher}, \& {Labrosse}}]{Brown2016}
{Brown}, S.~A., {Fletcher}, L., \& {Labrosse}, N. 2016, \aap, 596, A51, \dodoi{10.1051/0004-6361/201628390}

\bibitem[{{Chamberlin}(2016)}]{Chamberlin2016}
{Chamberlin}, P.~C. 2016, \solphys, 291, 1665, \dodoi{10.1007/s11207-016-0931-0}

\bibitem[{{Chen} {et~al.}(2022){Chen}, {Tian}, {Li}, {Wang}, {Lu}, {Xu}, {Hou}, \& {Wu}}]{Chen2022}
{Chen}, H., {Tian}, H., {Li}, H., {et~al.} 2022, \apj, 933, 92, \dodoi{10.3847/1538-4357/ac739b}

\bibitem[{{Cheng} {et~al.}(2019){Cheng}, {Wang}, {Liu}, {Zhou}, \& {Liu}}]{Cheng2019}
{Cheng}, Z., {Wang}, Y., {Liu}, R., {Zhou}, Z., \& {Liu}, K. 2019, \apj, 875, 93, \dodoi{10.3847/1538-4357/ab0f2d}

\bibitem[{{Dissauer} {et~al.}(2019){Dissauer}, {Veronig}, {Temmer}, \& {Podladchikova}}]{Dissauer2019}
{Dissauer}, K., {Veronig}, A.~M., {Temmer}, M., \& {Podladchikova}, T. 2019, \apj, 874, 123, \dodoi{10.3847/1538-4357/ab0962}

\bibitem[{{Dissauer} {et~al.}(2018){Dissauer}, {Veronig}, {Temmer}, {Podladchikova}, \& {Vanninathan}}]{Dissauer2018}
{Dissauer}, K., {Veronig}, A.~M., {Temmer}, M., {Podladchikova}, T., \& {Vanninathan}, K. 2018, \apj, 855, 137, \dodoi{10.3847/1538-4357/aaadb5}

\bibitem[{{Eparvier} {et~al.}(2009){Eparvier}, {Crotser}, {Jones}, {McClintock}, {Snow}, \& {Woods}}]{Eparvier2009}
{Eparvier}, F.~G., {Crotser}, D., {Jones}, A.~R., {et~al.} 2009, in Society of Photo-Optical Instrumentation Engineers (SPIE) Conference Series, Vol. 7438, Solar Physics and Space Weather Instrumentation III, ed. S.~{Fineschi} \& J.~A. {Fennelly}, 743804, \dodoi{10.1117/12.826445}

\bibitem[{{Gopalswamy} \& {Yashiro}(2013)}]{Gopalswamy2013}
{Gopalswamy}, N., \& {Yashiro}, S. 2013, \pasj, 65, S11, \dodoi{10.1093/pasj/65.sp1.S11}

\bibitem[{{Harra} \& {Sterling}(2001)}]{Harra2001}
{Harra}, L.~K., \& {Sterling}, A.~C. 2001, \apjl, 561, L215, \dodoi{10.1086/324767}

\bibitem[{{Hock}(2012)}]{Hock2012}
{Hock}, R.~A. 2012, PhD thesis, University of Colorado, Boulder

\bibitem[{{Hudson} {et~al.}(1996){Hudson}, {Lemen}, \& {Webb}}]{Hudson1996}
{Hudson}, H.~S., {Lemen}, J.~R., \& {Webb}, D.~F. 1996, in Astronomical Society of the Pacific Conference Series, Vol. 111, Astronomical Society of the Pacific Conference Series, ed. R.~D. {Bentley} \& J.~T. {Mariska}, 379--382

\bibitem[{{Hudson} {et~al.}(2011){Hudson}, {Woods}, {Chamberlin}, {Fletcher}, {Del Zanna}, {Didkovsky}, {Labrosse}, \& {Graham}}]{Hudson2011}
{Hudson}, H.~S., {Woods}, T.~N., {Chamberlin}, P.~C., {et~al.} 2011, \solphys, 273, 69, \dodoi{10.1007/s11207-011-9862-y}

\bibitem[{{Inoue} {et~al.}(2024){Inoue}, {Enoto}, {Namekata}, {Notsu}, {Honda}, {Maehara}, {Zhang}, {Lu}, {Uchida}, {Tsuru}, {Nogami}, \& {Shibata}}]{Inoue2024}
{Inoue}, S., {Enoto}, T., {Namekata}, K., {et~al.} 2024, \pasj, \dodoi{10.1093/pasj/psae001}

\bibitem[{{Jenkins} {et~al.}(2018){Jenkins}, {Long}, {van Driel-Gesztelyi}, \& {Carlyle}}]{Jenkins2018}
{Jenkins}, J.~M., {Long}, D.~M., {van Driel-Gesztelyi}, L., \& {Carlyle}, J. 2018, \solphys, 293, 7, \dodoi{10.1007/s11207-017-1224-y}

\bibitem[{{Jin} {et~al.}(2009){Jin}, {Ding}, {Chen}, {Fang}, \& {Imada}}]{Jin2009}
{Jin}, M., {Ding}, M.~D., {Chen}, P.~F., {Fang}, C., \& {Imada}, S. 2009, \apj, 702, 27, \dodoi{10.1088/0004-637X/702/1/27}

\bibitem[{{Lammer} {et~al.}(2007){Lammer}, {Lichtenegger}, {Kulikov}, {Grie{\ss}meier}, {Terada}, {Erkaev}, {Biernat}, {Khodachenko}, {Ribas}, {Penz}, \& {Selsis}}]{Lammer2007}
{Lammer}, H., {Lichtenegger}, H. I.~M., {Kulikov}, Y.~N., {et~al.} 2007, Astrobiology, 7, 185, \dodoi{10.1089/ast.2006.0128}

\bibitem[{{Leitzinger} \& {Odert}(2022)}]{Leitzinger2022}
{Leitzinger}, M., \& {Odert}, P. 2022, Serbian Astronomical Journal, 205, 1, \dodoi{10.2298/SAJ2205001L}

\bibitem[{{Leitzinger} {et~al.}(2011){Leitzinger}, {Odert}, {Ribas}, {Hanslmeier}, {Lammer}, {Khodachenko}, {Zaqarashvili}, \& {Rucker}}]{Leitzinger2011}
{Leitzinger}, M., {Odert}, P., {Ribas}, I., {et~al.} 2011, \aap, 536, A62, \dodoi{10.1051/0004-6361/201015985}

\bibitem[{{Lemen} {et~al.}(2012){Lemen}, {Title}, {Akin}, {Boerner}, {Chou}, {Drake}, {Duncan}, {Edwards}, {Friedlaender}, {Heyman}, {Hurlburt}, {Katz}, {Kushner}, {Levay}, {Lindgren}, {Mathur}, {McFeaters}, {Mitchell}, {Rehse}, {Schrijver}, {Springer}, {Stern}, {Tarbell}, {Wuelser}, {Wolfson}, {Yanari}, {Bookbinder}, {Cheimets}, {Caldwell}, {Deluca}, {Gates}, {Golub}, {Park}, {Podgorski}, {Bush}, {Scherrer}, {Gummin}, {Smith}, {Auker}, {Jerram}, {Pool}, {Soufli}, {Windt}, {Beardsley}, {Clapp}, {Lang}, \& {Waltham}}]{Lemen2012}
{Lemen}, J.~R., {Title}, A.~M., {Akin}, D.~J., {et~al.} 2012, \solphys, 275, 17, \dodoi{10.1007/s11207-011-9776-8}

\bibitem[{{Loyd} {et~al.}(2022){Loyd}, {Mason}, {Jin}, {Shkolnik}, {France}, {Youngblood}, {Villadsen}, {Schneider}, {Schneider}, {Llama}, {Ramiaramanantsoa}, \& {Richey-Yowell}}]{Loyd2022}
{Loyd}, R.~O.~P., {Mason}, J.~P., {Jin}, M., {et~al.} 2022, \apj, 936, 170, \dodoi{10.3847/1538-4357/ac80c1}

\bibitem[{{Lu} {et~al.}(2022){Lu}, {Tian}, {Zhang}, {Karoff}, {Chen}, {Shi}, {Hou}, {Chen}, {Xu}, {Wu}, {Cao}, \& {Wang}}]{Lu2022}
{Lu}, H.-p., {Tian}, H., {Zhang}, L.-y., {et~al.} 2022, \aap, 663, A140, \dodoi{10.1051/0004-6361/202142909}

\bibitem[{{Lu} {et~al.}(2023){Lu}, {Tian}, {Chen}, {Xu}, {Hou}, {Bai}, {Tan}, {Yang}, \& {Ren}}]{Lu2023}
{Lu}, H.-p., {Tian}, H., {Chen}, H.-c., {et~al.} 2023, \apj, 953, 68, \dodoi{10.3847/1538-4357/acd6a1}

\bibitem[{{Mason} {et~al.}(2014){Mason}, {Woods}, {Caspi}, {Thompson}, \& {Hock}}]{Mason2014}
{Mason}, J.~P., {Woods}, T.~N., {Caspi}, A., {Thompson}, B.~J., \& {Hock}, R.~A. 2014, \apj, 789, 61, \dodoi{10.1088/0004-637X/789/1/61}

\bibitem[{{Mason} {et~al.}(2016){Mason}, {Woods}, {Webb}, {Thompson}, {Colaninno}, \& {Vourlidas}}]{Mason2016}
{Mason}, J.~P., {Woods}, T.~N., {Webb}, D.~F., {et~al.} 2016, \apj, 830, 20, \dodoi{10.3847/0004-637X/830/1/20}

\bibitem[{{McCauley} {et~al.}(2015){McCauley}, {Su}, {Schanche}, {Evans}, {Su}, {McKillop}, \& {Reeves}}]{McCauley2015}
{McCauley}, P.~I., {Su}, Y.~N., {Schanche}, N., {et~al.} 2015, \solphys, 290, 1703, \dodoi{10.1007/s11207-015-0699-7}

\bibitem[{{Namekata} {et~al.}(2021){Namekata}, {Maehara}, {Honda}, {Notsu}, {Okamoto}, {Takahashi}, {Takayama}, {Ohshima}, {Saito}, {Katoh}, {Tozuka}, {Murata}, {Ogawa}, {Niwano}, {Adachi}, {Oeda}, {Shiraishi}, {Isogai}, {Seki}, {Ishii}, {Ichimoto}, {Nogami}, \& {Shibata}}]{Namekata2022}
{Namekata}, K., {Maehara}, H., {Honda}, S., {et~al.} 2021, Nature Astronomy, 6, 241, \dodoi{10.1038/s41550-021-01532-8}

\bibitem[{{Osten}(2023)}]{Osten2023}
{Osten}, R.~A. 2023, in Winds of Stars and Exoplanets, ed. A.~A. {Vidotto}, L.~{Fossati}, \& J.~S. {Vink}, Vol. 370, 25--36, \dodoi{10.1017/S1743921322003714}

\bibitem[{{Otsu} \& {Asai}(2024)}]{Otsu2024}
{Otsu}, T., \& {Asai}, A. 2024, \apj, 964, 75, \dodoi{10.3847/1538-4357/ad24ec}

\bibitem[{{Pesnell} {et~al.}(2012){Pesnell}, {Thompson}, \& {Chamberlin}}]{Pesnell2012}
{Pesnell}, W.~D., {Thompson}, B.~J., \& {Chamberlin}, P.~C. 2012, \solphys, 275, 3, \dodoi{10.1007/s11207-011-9841-3}

\bibitem[{{Reinard} \& {Biesecker}(2008)}]{Reinard2008}
{Reinard}, A.~A., \& {Biesecker}, D.~A. 2008, \apj, 674, 576, \dodoi{10.1086/525269}

\bibitem[{{Snow} {et~al.}(2009){Snow}, {McClintock}, {Crotser}, \& {Eparvier}}]{Snow2009}
{Snow}, M., {McClintock}, W.~E., {Crotser}, D., \& {Eparvier}, F.~G. 2009, in Society of Photo-Optical Instrumentation Engineers (SPIE) Conference Series, Vol. 7438, Solar Physics and Space Weather Instrumentation III, ed. S.~{Fineschi} \& J.~A. {Fennelly}, 743803, \dodoi{10.1117/12.828566}

\bibitem[{{Sterling} \& {Hudson}(1997)}]{Sterling1997}
{Sterling}, A.~C., \& {Hudson}, H.~S. 1997, \apjl, 491, L55, \dodoi{10.1086/311043}

\bibitem[{{Thompson} {et~al.}(1998){Thompson}, {Plunkett}, {Gurman}, {Newmark}, {St. Cyr}, \& {Michels}}]{Thompson1998}
{Thompson}, B.~J., {Plunkett}, S.~P., {Gurman}, J.~B., {et~al.} 1998, \grl, 25, 2465, \dodoi{10.1029/98GL50429}

\bibitem[{{Tian} {et~al.}(2012){Tian}, {McIntosh}, {Xia}, {He}, \& {Wang}}]{Tian2012}
{Tian}, H., {McIntosh}, S.~W., {Xia}, L., {He}, J., \& {Wang}, X. 2012, \apj, 748, 106, \dodoi{10.1088/0004-637X/748/2/106}

\bibitem[{{Vanninathan} {et~al.}(2018){Vanninathan}, {Veronig}, {Dissauer}, \& {Temmer}}]{Vanninathan2018}
{Vanninathan}, K., {Veronig}, A.~M., {Dissauer}, K., \& {Temmer}, M. 2018, \apj, 857, 62, \dodoi{10.3847/1538-4357/aab09a}

\bibitem[{{Veronig} {et~al.}(2019){Veronig}, {G{\"o}m{\"o}ry}, {Dissauer}, {Temmer}, \& {Vanninathan}}]{Veronig2019}
{Veronig}, A.~M., {G{\"o}m{\"o}ry}, P., {Dissauer}, K., {Temmer}, M., \& {Vanninathan}, K. 2019, \apj, 879, 85, \dodoi{10.3847/1538-4357/ab2712}

\bibitem[{{Veronig} {et~al.}(2021){Veronig}, {Odert}, {Leitzinger}, {Dissauer}, {Fleck}, \& {Hudson}}]{Veronig2021}
{Veronig}, A.~M., {Odert}, P., {Leitzinger}, M., {et~al.} 2021, Nature Astronomy, 5, 697, \dodoi{10.1038/s41550-021-01345-9}

\bibitem[{{Vida} {et~al.}(2019){Vida}, {Leitzinger}, {Kriskovics}, {Seli}, {Odert}, {Kov{\'a}cs}, {Korhonen}, \& {van Driel-Gesztelyi}}]{Vida2019}
{Vida}, K., {Leitzinger}, M., {Kriskovics}, L., {et~al.} 2019, \aap, 623, A49, \dodoi{10.1051/0004-6361/201834264}

\bibitem[{{Williams} {et~al.}(2013){Williams}, {Baker}, \& {van Driel-Gesztelyi}}]{Williams2013}
{Williams}, D.~R., {Baker}, D., \& {van Driel-Gesztelyi}, L. 2013, \apj, 764, 165, \dodoi{10.1088/0004-637X/764/2/165}

\bibitem[{{Woods} {et~al.}(2012){Woods}, {Eparvier}, {Hock}, {Jones}, {Woodraska}, {Judge}, {Didkovsky}, {Lean}, {Mariska}, {Warren}, {McMullin}, {Chamberlin}, {Berthiaume}, {Bailey}, {Fuller-Rowell}, {Sojka}, {Tobiska}, \& {Viereck}}]{Woods2012}
{Woods}, T.~N., {Eparvier}, F.~G., {Hock}, R., {et~al.} 2012, \solphys, 275, 115, \dodoi{10.1007/s11207-009-9487-6}

\bibitem[{{Xu} {et~al.}(2022){Xu}, {Tian}, {Hou}, {Yang}, {Gao}, \& {Bai}}]{Xu2022}
{Xu}, Y., {Tian}, H., {Hou}, Z., {et~al.} 2022, \apj, 931, 76, \dodoi{10.3847/1538-4357/ac69d5}

\bibitem[{{Zhukov} \& {Auch{\`e}re}(2004)}]{Zhukov2004}
{Zhukov}, A.~N., \& {Auch{\`e}re}, F. 2004, \aap, 427, 705, \dodoi{10.1051/0004-6361:20040351}

\end{thebibliography}

\end{document}